\documentstyle[pre,aps]{revtex}
\begin{document}
\draft
\font\sqi=cmssq8
\def\DR{\rm I\kern-1.45pt\rm R}
\def\DC{\kern2pt {\hbox{\sqi I}}\kern-4.2pt\rm C}
\def\DH{\rm I\kern-1.5pt\rm H\kern-1.5pt\rm I}
\def\bs{\mbox{\boldmath $\sigma$}}
\def\theequation{\arabic{equation}}
\twocolumn[\hsize\textwidth\columnwidth\hsize\csname
@twocolumnfalse\endcsname

\title{A note on $N=4$ supersymmetric mechanics on K\"ahler manifolds.}
\author{Stefano Bellucci$\;^1$ and Armen
Nersessian$\;^{1,2,3}$ }
\address{$\;^1$ INFN, Laboratori Nazionali di  Frascati, P.O. Box 13,
I-00044 Frascati, Italy\\
$\;^2$ JINR, Laboratory of Theoretical Physics,
 Dubna, 141980  Russia\\
$\;^3$ Yerevan State University, A.Manoogian, 1, Yerevan,
375025 Armenia}
\date{\today}
\maketitle
\begin{abstract}
\noindent
The geometric models of $N=4$ supersymmetric mechanics with
$(2d.2d)_{\DC}$-dimensional phase space are proposed,
which can be viewed as one-dimensional counterparts of two-dimensional
$N=2$ supersymmetric sigma-models by Alvarez-Gaum\'e and Freedman.
The related construction of supersymmetric mechanics
whose phase space is a K\"ahler supermanifold
is considered. Also, its relation with antisymplectic geometry is discussed.
\end{abstract}
\pacs{PACS number: 11.30.Pb}
]
\section{Introduction}
Supersymmetric mechanics attracts permanent interest
since its introduction \cite{witten}. However,
studies focussed mainly on the $N=2$ case, and the
most important case of $N=4$ mechanics did not receive enough
attention,
though some interesting observations were made about this subject:
let us mention that the most general
 $N=4, D=1,3$  supersymmetric mechanics described by real superfield
actions  were  studied in Refs. \cite{ikp,is} respectively, and those
in arbitrary $D$ in Ref.\cite{dpt};
in \cite{bp} $N=4, D=2$ supersymmetric mechanics described
by chiral superfield actions were considered;
the general study of supersymmetric mechanics with arbitrary
$N$ was performed recently in Ref.\cite{hull}.
In the Hamiltonian language classical supersymmetric mechanics
can be formulated
in terms of  superspace equipped with some
supersymplectic structure (and  corresponding non-degenerate
Poisson brackets). After quantization
the odd coordinates are replaced by the generators
of Clifford algebra. It is easy to verify that the
minimal dimension of phase superspace, which allows to describe
a $D-$dimensional supersymmetric mechanics
with {\sl nonzero} potential terms, is
$(2D.2D)$, while supersymmetry specifies
both the admissible sets of configuration spaces and potentials.

In the present  work we propose the $N=4$
supersymmetric one-dimensional
sigma-models (with and without central charge) on K\"ahler manifold
$(M_0, g_{a\bar b}dz^ad{\bar z}^{\bar b})$,
with $(2d.2d)_{\DC}$-dimensional  phase space
equipped with the symplectic structure
\begin{equation}
\begin{array}{c}
\Omega=\omega_0-i\partial{\bar\partial}{\bf g}=\\
=d\pi_a\wedge dz^a+ d{\bar\pi}_a\wedge d{\bar z}^a+\\
+R_{a{\bar b}c\bar d}\eta^a_i\bar\eta^b_i dz^a\wedge d{\bar z}^b+
g_{a\bar b}D\eta^a_i\wedge{D{\bar\eta}^b_i}
\end{array}
\label{ss}\end{equation}
where
\begin{equation}
{\bf g}=ig_{a\bar b}\eta^a\sigma_0{\bar\eta}^b,\quad
D\eta^a_i=d\eta^a_i+\Gamma^a_{bc}\eta^a_i dz^a,\quad i=1,2
\end{equation}
while $\Gamma^a_{bc},\; R_{a\bar b c\bar d}$ are
respectively the connection and curvature of the K\"ahler structure.
The odd coordinates $\eta^a_i$ belong to the external algebra
$\Lambda(M_0)$, i.e. they transform as  $dz^a$.
This symplectic structure becomes canonical
in the coordinates $(p_a,\chi^k)$
\begin{equation}
\begin{array}{c}
p_a=\pi_a-\frac{i}{2} \partial_a{\bf g},
\quad\chi^m_i={\rm e}^m_b\eta^b_i:\\
\Omega=dp_a\wedge d z^a +d{\bar p}_{\bar a}\wedge d{\bar z}^{\bar a}
+d\chi^m_i\wedge d{\bar\chi}^{\bar m}_i,
\end{array}
\label{canonical}\end{equation}
where ${\rm e}^m_a$ are the einbeins of the  K\"ahler structure:
${\rm e}^m_a\delta_{m\bar m}{\bar{\rm e}}^{\bar m}_{\bar b}=g_{a\bar b}.$
So, to quantize this model, one chooses
$$ {\hat p}_a=-i\frac{\partial}{\partial z^a},\quad
 {\hat{\bar  p}}_{\bar a}=-i\frac{\partial}{\partial \bar z^{\bar a}},\quad
[{\hat\chi}^m_i,{\hat{\bar\chi}}^{\bar n}_j]_+
=\delta^{m\bar n}\delta_{ij}.
$$
We  restrict ourselves by   the supersymmetric mechanics
whose  supercharges  are {\it linear}  in the
Grassmann variables $\eta^a_i$, $\bar\eta^{\bar a}_i$.
These systems can be obtained by dimensional
reduction from  $N=2$ supersymmetric $(1+1)-$dimensional sigma-models
 by Alvarez-Gaum\'e and Freedman \cite{agf};
in the simplest case of $d=1$  and in the absence of central
charge these systems coincide with the $N=4$ supersymmetric mechanics
described by the chiral superfield action \cite{bp}.
The constructed systems are connected also with the $N=4$
supersymmetric mechanics describing the low-energy
dynamics of monopoles and dyons
in  $N=2,4$ super-Yang-Mills theories \cite{gk}.

We also propose the related construction of $N=2$ supersymmetric
mechanics  whose  phase superspace is the external algebra
of an arbitrary K\"ahler manifold. Under the additional
assumption that the base manifold is a hyper-K\"ahler one, this
system should get the $N=4$ supersymmetry. The relation
of this system with antisymplectic geometry is discussed.

\section{Sigma-model with standard  $N=4$ SUSY.}
Let us consider a one-dimensional
supersymmetric sigma-model on an arbitrary Riemann manifold
 $(M_0, g_{\mu\nu}(x)dx^\mu dx^\nu)$,  with  $(2D.2D)-$dimensional
 phase superspace equipped with a supersymplectic structure
\begin{equation}
\begin{array}{c}
\Omega=d\left(p_\mu dx^\mu +
\theta^\mu_i g_{\mu\nu}D\theta^\nu_i
\right)= \\
=dp_\mu\wedge dx^\mu+
\frac 12 R_{\mu\nu\lambda\rho}\theta^\mu_i\theta^\nu_i
 dx^\lambda\wedge dx^\rho\\
+g_{\mu\nu}D\theta^\mu_i\wedge D\theta^\nu_i,
\end{array}
\end{equation}
where
$$
D\theta^\mu_i=
d\theta^\mu_i +
\Gamma^\nu_{\rho\lambda}\theta^\rho_i dx^\lambda,
$$
and  $\Gamma^\mu_{\nu\lambda}$, $R_{\mu\nu\lambda\rho}$
 are respectively the Cristoffel symbols and curvature tensor of
underlying metric $g_{\mu\nu}dx^\mu dx^\nu$.
On this phase superspace one can formulate the  one-dimensional
$N=2$ supersymmetric sigma-model with
supercharges linear on Grassmann variables, viz
\begin{equation}
\begin{array}{c}
Q_1=p_\mu\theta^\mu_1+ U_{,\mu}(x)\theta^\mu_2,
\quad Q_1=p_\mu\theta^\mu_2- U_{,\mu}(x)\theta^\mu_1,\\
{\cal H}=\frac 12 g^{\mu\nu}\left(p_\mu p_\nu+ U_{,\mu}U_{,\nu}\right)+
U_{\mu;\nu}\theta^\mu_1\theta^\nu_2 +\\
+R_{\mu\nu\lambda\rho}\theta^\mu_1\theta^\nu_2
\theta^\lambda_1\theta^\rho_2:\\
\{Q_i,Q_j\}=2\delta_{ij}{\cal H},\quad
\{Q_i,{\cal H}\}=0, i=1,2.
\end{array}
\end{equation}
One can also introduce a specific constant of motion
(``fermionic number'')
\begin{equation}
\begin{array}{c}
{\cal F}=g_{\mu\nu}\theta^\mu_1\theta^\nu_2\;:\;
\{{\cal F},Q_i\}=\epsilon_{ij}Q_j\;
 \{{\cal F},{\cal H}\}=0.
\end{array}
\end{equation}
To get the $N=4$ supersymmetric one-dimensional
sigma-model mechanics, one should require that
the target space $M_0$ is a K\"ahler manifold
$(M_0, g_{a\bar b}dz^ad{\bar z}^{\bar b})$,
$g_{a\bar b}=\partial^2 K(z,\bar z)/\partial z^a\partial{\bar z}^b$
(this restriction follows also from the
considerations of superfield actions:
indeed, the $N-$extended supersymmetric
mechanics obtained from the
action depending on $D$  real superfields,
have a $(2D.ND)_{\DR}$-dimensional symplectic manifold,
whereas those obtained from the action depending on  $d$
chiral superfields have a $(2d.Nd/2)_{\DC}-$dimensional
phase space, with the configuration space being a
$2d-$dimensional K\"ahler manifold).
In that case the phase superspace
can be equipped  by the supersymplectic structure (\ref{ss}).
The corresponding Poisson brackets are defined
by the following non-zero
relations (and their complex-conjugates):
$$
\begin{array}{c}
\{\pi_a, z^b\}=\delta^b_a,\quad
\{\pi_a,\eta^b_i\}=-\Gamma^b_{ac}\eta^c_i,\\
\{\pi_a,\bar\pi_b\}=-R_{a\bar b c\bar d}\eta^c_k{\bar\eta}^d_k,\quad
\{\eta^a_i, \bar\eta^b_j\}=g^{a\bar b}\delta_{ij}.
\end{array}
$$

To construct on this phase superspace
the Hamiltonian mechanics  with  standard $N=4$ supersymmetry algebra
 \begin{equation}
\begin{array}{c}
\{Q^+_i,Q^-_j\}=\delta_{ij}{\cal H},\\
\{Q^\pm_i,Q^\pm_j\}=\{Q^\pm_i, {\cal H}\}=0,\quad i=1, 2,
\end{array}
\label{4sualg}\end{equation}
let us choose the  supercharges  given by the functions
 \begin{equation}
Q^+_1=\pi_a\eta^a_1+ iU_{\bar a}{\bar \eta}^{\bar a}_2,\quad
Q^+_2=\pi_a\eta^a_2- iU_{\bar a}{\bar \eta}^{\bar a}_1.
\label{4SUSY}\end{equation}
Then, calculating the commutators (Poisson brackets) of these
functions,
we get that the supercharges (\ref{4SUSY})  belong
to the superalgebra (\ref{4sualg})
when the functions $U_a, {\bar U}_{\bar a}$
are  of the form
\begin{equation}
U_a(z)=\frac{\partial U(z)}{\partial z^a},\quad
{\bar U}_{\bar a}(\bar z )=\frac{\partial {\bar U}({\bar z})}{\partial {\bar z}^a},
\end{equation}
while   the Hamiltonian reads
\begin{equation}
\begin{array}{c}
{\cal H}=g^{a{\bar b}}(\pi_a{\bar\pi}_b+
{U}_a{\bar U}_{\bar b})-iU_{a;b}\eta^a_1\eta^{b}_2
+i{\bar U}_{\bar a;\bar b}{\bar\eta}^{\bar a}_1{\bar\eta}^{\bar b}_2-\\
-R_{a\bar b c\bar d}\eta^a_1\bar\eta^b_1\eta^a_2\bar\eta^d_2,
\end{array}
\label{4SUHam}\end{equation}
 where
$ U_{a;b}\equiv \partial_a\partial_b U-\Gamma^c_{ab}\partial_cU$.

The constant of motion  counting the number of fermions, reads:
\begin{equation}
{\cal F}=ig_{a\bar b}\eta^a\sigma_3{\bar\eta}^{\bar b}:
\quad
 \{Q^\pm_i, {\cal F}\}=\pm i Q^\pm_i, \; \{{\cal H}, {\cal F}\}=0.
\label{f}\end{equation}
Performing the Legendre transformation
one gets the  Lagrangian of the system
\begin{equation}
\begin{array}{c}
{\cal L}=
g_{a{\bar b}}{\dot z}^a{\dot{\bar z}^{\bar b}}-
\frac 12 \eta^a_kg_{a\bar b}\frac{D{\bar\eta}^{\bar b}_k}{d\tau}+
\frac 12 \frac{D\eta^{a}_k}{d\tau}
g_{a\bar b}\bar\eta^{\bar b}-\\
-g^{a{\bar b}}{U}_a{\bar U}_{\bar b}
+iU_{a;b}\eta^a_1\eta^{b}_2
-i{\bar U}_{\bar a;\bar b}{\bar\eta}^{\bar a}_1{\bar\eta}^{\bar b}_2+\\
+R_{a\bar b c\bar d}\eta^a_1\bar\eta^b_1\eta^c_2\bar\eta^d_2.
\end{array}
\end{equation}
The supersymmetry transformations of the Lagrangian are of the form
\begin{equation}
\begin{array}{c}
\delta^+_i z^a=\epsilon\eta^a_i,\\
\delta^+_i \eta^a_j=\epsilon
\left(i\epsilon_{ij}{\bar U}_{\bar b}g^{\bar b a}+
\Gamma^a_{bc}\eta^b_i\eta^c_j\right);\\
 \delta^-_i z^a=0,\\
\delta^-_i\eta^a_j=\epsilon\delta_{ij}{\dot z}^a
\end{array}
\end{equation}
where $\epsilon$ is an odd parameter: $p(\epsilon)=1$.\\
So, we get  the action for one-dimensional sigma-model
 with four exact real supersymmetries. It can be straightly obtained
 by the dimensional reduction of $N=2$ supersymmetric
$(1+1)$ dimensional sigma-model by
 Alvarez-Gaum\'e and Freedman \cite{agf}
(the mechanical counterpart of this system without
potential term was constructed in \cite{macfarlane}).
Notice that the above-presented $N=4$ SUSY
mechanics for the simplest case, i.e. $d=1$,
was obtained by Berezovoy and Pashnev \cite{bp}
from the chiral superfield action
 \begin{equation}
 {\cal S}=\frac 12\int K(\Phi,\bar\Phi)
+2\int U(\Phi)+2\int{\bar U}(\bar\Phi)
\end{equation}
where $\Phi$ is  chiral superfield.
It seems to be obvious that a similar action depending on $d$
chiral superfields will generate the above-presented $N=4$ SUSY
mechanics.
\section{$N=4$ sigma-model with central charge}
Let us consider a generalization of above system,
which possesses $N=4$  supersymmetry with
central charge
\begin{equation}
\begin{array}{c}
\{\Theta^+_i,\Theta^-_j\}=
\delta_{ij}{\cal H}+{\cal Z}\sigma^3_{i{j}},\quad
\{\Theta^\pm_i,\Theta^\pm_j\}=0,
\\
\{{\cal Z}, {\cal H}\}= \{{\cal Z},\Theta^\pm_k\}=0.
\end{array}
\label{csa}\end{equation}
For this purpose one introduces
the supercharges
\begin{equation}
\begin{array}{c}
 \Theta^+_1=\left(\pi_a+iG_{,a}(z,\bar z)\right)\eta^a_1 +
i {\bar U}_{,\bar a}({\bar z}){\bar \eta}^{\bar a}_2,\\
 \Theta^+_2=\left(\pi_a-iG_{,a}(z,\bar z)\right)\eta^a_2 -
i {\bar U}_{,{\bar a}}({\bar z}){\bar \eta}^{\bar a}_1,
\end{array}
\end{equation}
where the real function $G(z,\bar z)$  obeys the conditions
\begin{equation}
\partial_a\partial_b G
+\Gamma^c_{ab}\partial_c G=0,\quad
{ G}_{,a}(z,\bar z)g^{a\bar b}{\partial}_{\bar b}{\bar U}({\bar z})=0.
\label{killing}\end{equation}
The first equation in (\ref{killing}) is nothing
but the Killing equation
of  the underlying K\"ahler structure
(let us remind, that the isometries of the K\"ahler structure are
Hamiltonian holomorphic vector fields) given by the vector
\begin{equation}
{\bf G}=G^a(z)\partial_a+{\bar G}^a({\bar z}){\bar\partial}_a,\quad
G^a=ig^{a\bar b}{\bar \partial}_b G.
\end{equation}
The second equation means that the vector field ${\bf G}$
leaves the holomorphic function invariant:
$${\cal L}_{\bf G}U=0\;\Rightarrow \;G^a(z)U_a(z)=0.$$

Calculating the Poisson brackets of these supercharges,
we get explicit
expressions for the Hamiltonian
\begin{equation}
\begin{array}{c}
{\cal H}\equiv
g^{a{\bar b}}\left(\pi_a{\bar\pi}_{\bar b}+ G_{,a}G_{{\bar b}}
+{ U}_{,a}{\bar U}_{,\bar b}\right)- \\
-iU_{a;b}\eta^a_1\eta^{b}_2 +
i{\bar U}_{\bar a;\bar b}{\bar\eta}^{\bar a}_1{\bar\eta}^{\bar b}_2
+\frac 12 G_{a\bar b}(\eta^a_k\bar\eta^{\bar b}_k) -\\
-R_{a\bar b c\bar d}\eta^a_1\bar\eta^b_1\eta^c_2\bar\eta^d_2 
\end{array}
\end{equation}
and the central charge
\begin{equation}
\begin{array}{c}
{{\cal Z}}=i(G^a\pi_a+G^{\bar a}{\bar\pi}_{\bar a})+\frac 12
\partial_a{\bar\partial}_{\bar b}G(\eta^a{\sigma_3}\bar\eta^{\bar b}).
\end{array}
\label{z}\end{equation}
It can be checked by a straightforward calculation that
the function ${\cal Z}$ indeed belongs to the center of
the superalgebra (\ref{csa}).
The scalar part of each  phase
with  standard $N=2$ supersymmetry can be interpreted
as a particle  moving on the K\"ahler  manifold
in the presence of an external magnetic field with strength
$F=iG_{a\bar b}dz^a\wedge d{\bar z}^{\bar b}$ and in the
potential field $U_{,a}(z)g^{a\bar b}{\bar U}_{,\bar b}(\bar z)$.

The Lagrangian of the system is of the form
\begin{equation}
\begin{array}{c}
{\cal L}=
g_{a{\bar b}}\left({\dot z}^a{\dot{\bar z}}^{b}+
\frac 12 \eta^a_k\frac{D{\bar\eta}^{\bar b}_k}{d\tau}+
\frac 12 \frac{D\eta^{a}_k}{d\tau}\bar\eta^{\bar b}\right)-\\
-g^{a{\bar b}}(G_aG_{\bar b}+{U}_a{\bar U}_{\bar b})+\\
+iU_{a;b}\eta^a_1\eta^{b}_2-
i{\bar U}_{\bar a;\bar b}{\bar\eta}^{\bar a}_1{\bar\eta}^{\bar b}_2+
R_{a\bar b c\bar d}\eta^a_1\bar\eta^b_1\eta^a_2\bar\eta^d_2.
\end{array}
\end{equation}
The supersymmetry transformations read
\begin{equation}
\begin{array}{c}
\delta^+_i z^a=\epsilon\eta^a_i,\\
\delta^+_i \eta^a_j=
\epsilon\left(i\epsilon_{ij}{\bar U}_{\bar b}g^{\bar b a}
+ \Gamma^a_{bc}\eta^b_i\eta^c_j\right),\\
\delta^-_i z^a=0,\\
\delta^-_i\eta^a_j=\epsilon(\delta_{ij}{\dot z}^a-\epsilon_{ij}G^a).
\end{array}
\end{equation}
Assuming that $(M_0, g_{a\bar b}dz^a d{\bar z}^b)$ is the
hyper-K\"ahler metric and that $U(z)+{\bar U}({\bar z})$
is a tri-holomorphic function while the function
$G(z,\bar z)$ defines a tri-holomorphic Killing vector,
one should get the $N=8$ supersymmetric one-dimensional sigma-model.
In that case instead of the
phase with standard $N=2$ SUSY arising in the
K\"ahler case, we shall get the phase with standard $N=4$ SUSY.
The latter system can be viewed as a particular
case of $N=4$ SUSY mechanics describing the
low-energy dynamics of monopoles and dyons in
$N=2,4$ super-Yang-Mills theory\cite{gk}.
Notice that, in contrast to the $N=4$ mechanics suggested
in the mentioned
papers, in the above-proposed (hypothetic) construction
also the four hidden supersymmetries could be explicitly written.
We wish to consider this $N=8$ supersymmetric mechanics,
as well as its
application to the
solutions of super-Yang-Mills theory, in a forthcoming paper.

\section{The related construction}
Let us  consider a supersymmetric mechanics
 whose phase superspace is the external
algebra of the K\"ahler manifold $\Lambda({M})$, where
$\left({M},\; g_{A\bar B}(z,\bar z)dz^A d{\bar z}^{\bar B}\right)$
 plays the role of the phase space
of the underlying Hamiltonian mechanics.
The phase superspace is a
$(D.D)_{\DC}-$ dimensional K\"ahler supermanifold
 equipped by  the super-K\"ahler
 structure \cite{knkahler}
\begin{equation}
\begin{array}{c}
\Omega=i\partial \bar\partial
\left(K(z,\bar z)-ig_{A\bar B}\theta^A{\bar\theta}^{\bar B}
\right)=\\
=i(g_{A\bar B}-i
R_{A\bar BC\bar D}\theta^C\bar\theta^{\bar D})dz^A\wedge
 d{\bar z}^{\bar B}
\\+g_{A\bar B}D\theta^A\wedge D{\bar\theta}^{\bar B},
\end{array}
\label{ssg}\end{equation}
where
$D\theta^A=d\theta^A +
\Gamma^A_{BC}\theta^B dz^C$, and
 $\Gamma^A_{BC}$, $R_{A\bar BC\bar D}$ are respectively
the Cristoffel symbols and curvature tensor of the underlying
K\"ahler metrics $g_{A\bar B}=\partial_A\partial_{\bar B}K(z,\bar z)$.

The corresponding Poisson bracket can be presented in the form
\begin{equation}
\begin{array}{c}
\{\quad,\quad\}=i{\tilde g}^{A\bar B}\nabla_A
\wedge{\bar\nabla}_{\bar B}+g^{A\bar B}\frac{\partial}{\partial \theta^A}
\wedge\frac{\partial}{\partial{\bar\theta}^{\bar B}}\; ,\;
\end{array}
\end{equation}
where
$$
\nabla_A=\frac{\partial}{\partial z^A}-
\Gamma^C_{AB}\theta^B\frac{\partial}{\partial\theta^C}
$$
and
$$
{\tilde g}^{-1}_{A\bar B}=(g_{A\bar B}-
 iR_{A\bar BC\bar D}\theta^C\bar\theta^{\bar D}).
$$
On this phase superspace one can immediately construct
the mechanics  with $N=2$ supersymmetry
 \begin{equation}
\{Q_+,Q_-\}={\tilde{\cal H}},\quad\{Q_\pm,Q_\pm\}=\{Q_\pm, {\tilde{\cal H}}\}=0,
\label{2psualg}\end{equation}
 given by the supercharges
\begin{equation}
\begin{array}{c}
Q_+=\partial_A H(z, \bar z)\theta^A,\quad
Q_-=\partial_{\bar A}H(z, \bar z){\bar\theta}^{\bar A}
\end{array}
\end{equation}
where $H(z,\bar z)$
is the Killing potential of the underlying K\"ahler structure,
$$ {\partial_A\partial_B}H-\Gamma^C_{AB}\partial_C H=0,\quad
V^A(z)=ig^{A\bar B}\partial_{\bar B}H(z,\bar z).
$$
The Hamiltonian of the system reads
\begin{equation}
\begin{array}{c}
{\tilde{\cal H}}=g_{A\bar B}V^A{\bar V}^{\bar B}+
 iV^A_{,C}g_{A\bar B}{\bar V}^{\bar B}_{,\bar D}
\theta^C{\bar\theta}^{\bar D}-\\
-R_{A{\bar B} C{\bar D}}V^A_{,C}{\bar V}^{\bar B}_{;\bar D}
\theta^A\theta^C
{\bar\theta}^{\bar B}{\bar \theta}^{\bar D},
\end{array}
\end{equation}
while the supersymmetry transformations are given by the vector fields
$\delta^\pm\equiv \{Q^\pm,\; \}$,
\begin{equation}
\begin{array}{c}
\delta^-=-iV^A(z)\frac{\partial}{\partial \theta^A}-i
V^A_{;C}\theta^C{\cal N}^{D}_A\nabla_D,
\end{array}
\end{equation}
  where
$$
\begin{array}{c}
 \left({\cal N}^{-1}\right)^A_B\equiv
\delta^A_B-iR^A_{BC\bar D}\theta^C{\bar\theta}^{\bar D}.
\end{array}
$$
Requiring that ${M}$ be a hyper-K\"ahler manifold, we can
double the number
of supercharges and get a $N=4$ supersymmetric mechanics.
In that case the Killing potential should generate a
tri-holomorphic vector field.

The phase space of the system under consideration can be equipped,
in addition to the Poisson bracket corresponding to (\ref{ssg}),
with the  antibracket (odd Poisson bracket)
 associated with the odd K\"ahler structure
$\Omega_1=i\partial{\bar\partial}K_1,$
where $K_1={\rm e}^{i\alpha}{\partial_A K(z,\bar z)}\theta^A
+{\rm e}^{-i\alpha}{\partial_{\bar A}} K(z,\bar z)\bar\theta^{\bar A}$,
$\alpha=0,\pi/2$,
\begin{equation}
\{\quad,\quad\}_1={\rm e}^{-i\alpha}{g^{{\bar A} B}}\nabla_{\bar A}\wedge
\frac{\partial}{\partial\theta^B} +c.c.\quad.
\end{equation}
It is easy to observe that the following equality holds\cite{knkahler}
\begin{equation}
{\bf L}\equiv\{{\tilde {\cal Z}}, \;\;\}=\{ Q, \;\;\}_1,
\label{volk}\end{equation}
where
\begin{equation}
\begin{array}{c}
 {\tilde{\cal Z}}\equiv H(z,\bar z)+
i\partial_A\partial_{\bar B}H(z,\bar z)\theta^A{\bar\theta}^{\bar B}\\
 Q={\rm e}^{i\alpha} Q_+  +{\rm e}^{-i\alpha}Q_- .
\end{array}
\end{equation}
Then, after obvious algebraic manipulation with the Jacobi identities,
one gets the following relations:
\begin{equation}
\{{\tilde{\cal Z}}, {\tilde{\cal H}}\}= \{{\tilde{\cal Z}},Q_\pm\}=0.
\end{equation}
Hence, the function ${\tilde{\cal Z}}$ is a constant of motion,
which belongs to the center of the superalgebra
defined by $Q_\pm, {\cal H}, {\tilde{\cal Z}}$.
One can also introduce another constant of motion, i.e.
the ``fermionic number"
\begin{equation}
{\tilde {\cal F}}=ig_{AB}\theta^A{\bar\theta}^{\bar B}:
\;
 \{Q_\pm, {\tilde{\cal F}}\}=\pm i Q_\pm, \;
\{{\tilde{\cal H}}, {\tilde{\cal F}}\}=0.
\end{equation}
Notice that the supermanifolds provided by the even and odd
 symplectic (and K\"ahler) structures, and particularly the equation
(\ref{volk}), were studied in the context of the problem of
the description of  supersymmetric mechanics
in terms of antibrackets \cite{knkahler,knj}
(this problem was suggested in \cite{vpst}).
Later  this structure was
 found to be useful in  equivariant cohomology,
 e.g.  for the  construction of equivariant
characteristic classes and derivation of
localization formulae \cite{ec}, since the vector field
(\ref{volk}) can be identified with the
Lie derivative along the vector
field generated by $H$, while the vector fields
$\{{\tilde{\cal F}},\;\;\}_1$, $\{H,\;\;\}_1$
corresponds to external differential
and operator of inner  product.
Hence, $\{H+{\tilde{\cal F}},H+{\tilde{\cal F}} \}_1=2Q$,
and $H+{\tilde{\cal F}}$ defines an equivariant Chern class;
the Lie derivative of the even symplectic structure (\ref{ssg}) along
the vector field $\{H+{\tilde{\cal F}},\;\;\}_1$ yields the equivariant
even pre-symplectic structure, generating equivariant Euler classes of
the underlying K\"ahler manifold.

Notice also, that the antibrackets are the basic object of the
Batalin-Vilkovisky formalism \cite{bv}, while
the pair of antibrackets, corresponding to the $\alpha=0,\pi/2$,
 together with corresponding
 nilpotent  vector fields $\{{\cal F},\;\;\}_1$
 and the associated $\Delta-$operators, form the
 ``triplectic algebra" underlying the BRST-antiBRST-invariant
extension of Batalin-Vilkovisky formalism \cite{blt}.

\section{Acknowledgments}The  authors are grateful to
S. Krivonos for  numerous illuminating discussions,
and to E. Ivanov and A. Pashnev
for  useful  comments.
A.N. thanks INFN for financial support and kind hospitality
during his stay in Frascati,
within the  framework of a INFN-JINR agreement.


\begin{thebibliography}{99}
\bibitem{witten}E. Witten, Nucl. Phys. {\bf B188} (1981), 513;
{\it ibid.} {\bf B202} (1982), 253.
\bibitem{ikp}E.A. Ivanov, S.O. Krivonos, A.I. Pashnev,
Class. Quant. Grav.
{\bf 8} (1991), 19.
\bibitem{is}E.A. Ivanov, A.V. Smilga, Phys.~Lett. {\bf B257} (1991), 79;

 V.P. Berezovoy, A.I. Pashnev, Class. Quant. Grav.
{\bf 8} (1991), 2141.
\bibitem{dpt}E.E.~Donets, A.~Pashnev, J.J.~Rosales, M.~Tsulaia,
Phys.~Rev. {\bf D61} (2000), 043512.
\bibitem{bp}V.~Berezovoy, A.~Pashnev, Class.~Quant.~Grav. {\bf 13} (1996),
1699.
\bibitem{hull}C.M.~Hull, {\sl The   Geometry of Supersymmetric
Quantum Mechanics}, {\tt hep-th/9910028}.
\bibitem{gk} D.~Bak, K.~Lee, P.~Yi, Phys.~Rev. {\bf D61} (2000), 045003;
{\it ibid.} {\bf D62} (2000), 025009;

D.~Bak, C.~Lee, K.~Lee, P.~Yi, Phys.~Rev. {\bf D61} (2000), 025001;

J.~Gauntlett, N.~Kim, J.~Park, P.~Yi, Phys.~Rev. {\bf D61} (2000), 125012;

J.~Gauntlett, C.~Kim, K.~Lee, P.~Yi, {\tt hep-th/0008031}.
\bibitem{agf}L.~Alvarez-Gaum\'e, D.~Freedman,
Comm. Math. Phys. {\bf 91} (1983), 87; for the quantum connection of
sigma-models to strings see e.g. \cite{io} and references therein.
\bibitem{io}S. Bellucci, Z. Phys. C36 (1987), 229;
{\it ibid.} Prog. Theor. Phys. 79 (1988), 1288;
{\it ibid.} Z. Phys. C41 (1989), 631;
{\it ibid.} Phys. Lett. B227 (1989), 61; 
{\it ibid.} Mod. Phys. Lett. A5 (1990), 2253.
\bibitem{macfarlane}H.W.~Braden, A.J.~Macfarlane,
J.~Phys. {\bf A18} (1985), 2955.

\bibitem{knkahler}A.P.~Nersessian, Theor.~Math.~Phys. {\bf 96} (1993), 866.
\bibitem{knj}O.M.~Khudaverdian, J.~Math.~Phys. {\bf 32} (1991), 1934.

O.M.~Khudaverdian, A.P.~Nersessian, J.~Math.~Phys.{\bf 32} (1991), 1938;
{\it ibid.} {\bf 34} (1993), 5533.
\bibitem{vpst}D.V.~Volkov, A.~I.~Pashnev, V.A.~Soroka, V.I.~Tkach,
 JETP Lett. {\bf 44} (1986), 70.
\bibitem{ec} A.P. Nersessian,
JETP Lett. {\bf 58} (1993), 66;
Lecture Notes in Physics {\bf 524}, 90 (hep-th/9811110).
\bibitem{bv}I.A.~Batalin, G.A.~Vilkovisky,
 Phys.~Lett. {\bf B102} (1981), 27; Phys.~Rev. {\bf D28} (1983), 2567.

\bibitem{blt}I.A.~Batalin, P.M.~Lavrov, I.V.~Tyutin,
J.~Math.~Phys. {\bf 31} (1990), 1487;
{\it ibid.} {\bf 32} (1990), 532;

I.A.~Batalin, R.~Martnelius, A.~Semikhatov, Nucl.~Phys.
{\bf B 446} (1995), 249;

A.~Nersessian, P.H.~Damgaard, Phys. Lett. {\bf B355} (1995), 150;

I.A.~Batalin, R.~Marnelius, Nucl.~Phys. {\bf B465} (1996), 521.
\end{thebibliography}
\end{document}